\documentclass[aps,twocolumn,showpacs]{revtex4} 
\usepackage{graphicx} 

\begin{document} 

\title{Causality and Cirel'son bounds} 
\author{H. Buhrman} 
\email{buhrman@cwi.nl} 
\affiliation{C.W.I. and University of Amsterdam, P.O. Box 94079, 
1090 GB Amsterdam, The Netherlands} 
\author{S. Massar} 
\email{smassar@ulb.ac.be} 
\affiliation{Laboratoire d'Information Quantique and QUIC, 
{C.P.} 
165/59, Universit\'{e} Libre de Bruxelles, Avenue F. D. Roosevelt 
50, 1050 Bruxelles, Belgium} 

\begin{abstract} 
We study the properties of no signalling 
correlations that cannot 
be reproduced by local measurements on entangled quantum 
states. We say that such correlations violate Cirel'son bounds. We 
show that if these correlations are obtained by some  reversible unitary
quantum evolution $U$, then  $U$ cannot be written in the product form $U_A 
\otimes U_B$. This implies that $U$ can be used for 
signalling and for entanglement generation. This result is completely 
general and in fact can be viewed as a characterisation of 
Cirel'son bounds. We then show how this result can be used as a tool to study 
Cirel'son bounds and we illustrate this by rederiving the Cirel'son bound 
of $2\sqrt{2}$ for the CHSH inequality, and by deriving a new 
Cirel'son bound for qutrits.
\end{abstract} 

\maketitle 

\section{Introduction} 

Quantum non locality 
arises when measurements are carried out on two entangled 
particles in spatially separated regions. 
As first shown by Bell \cite{B}, the correlations obtained in such an 
experiment 
cannot be reproduced using classical local models, often called ``local 
hidden variable'' (lhv) models. 
Non locality is a fascinating 
chapter of physics and has attracted much attention since its 
discovery because it relates two fundamental aspects of 
nature, special relativity and quantum mechanics. 

Because information cannot travel faster than the speed of light, and 
since 
the measurements are carried out in 
spatially separated regions, 
such a setup cannot be used to transmit information from one site to 
the other. This constraint is expressed formally by the ``no signalling 
conditions'' we describe explicitly below.

The amount of non locality present in such correlations can be studied 
quantitatively by introducing the concept of "Bell expression". This 
is an expression which is bounded by a certain value for lhv models, 
but can exceed this value in the case of quantum correlations. 
We illustrate this by 
the well known 
Clauser-Horn-Shimony-Holt (CHSH) expression\cite{CHSH}. The CHSH expression is 
bounded by 2 for lhv models. But local measurements carried out on 
entangled quantum systems can reach the value of $2\sqrt{2}$. 
Cirel'son\cite{C} showed that this is the maximum value attainable by 
local measurements on entangled quantum systems. 

However there exist correlations which obey the no signalling conditions and 
for which the CHSH expression reaches the value of 4 (this is the 
maximum value consistent with positivity of the probabilities). Popescu and 
Rohrlich\cite{PR} were the first to study these maximally non local 
correlations as objects of interest in their own right. This lead them to 
raise a fundamental question: why isn't nature maximally non 
local? Providing an answer to this question would presumably deepen 
our understanding of relativity and quantum mechanics.

The CHSH expression is not the only way to study the non locality of 
quantum correlations. One can generalise it in many ways, for instance 
by changing the quantum states, by changing the measurements and in 
particular letting the measurements have more settings and/or more 
outcomes, 
or by having more 
entangled parties. In all cases the non locality of the quantum 
correlations can be studied using Bell inequalities which generalise the 
CHSH inequality. For any such Bell inequality there will be three, 
generally distinct, 
characteristic values: the maximal value attainable by lhv models, the 
maximal value attainable by local measurements on entangled quantum 
state, and the maximal value compatible with the no signalling 
conditions. In the present work, by analogy with the CHSH expression, 
we will call ``Cirel'son's bound'' the maximal value of any Bell 
expression 
attainable by local measurements on entangled quantum 
states. 

The boundary of the space of correlations which can be attained using 
a given model (lhv theory, quantum mechanics) presents a fundamental 
character. But it is very difficult to 
find this boundary in general. In the case of the lhv models, finding 
the boundary can be formulated as the mathematical problem of finding 
the facets of a polytope of which the vertices are known\cite{C2,MNL}. 

In the case of the boundary of the space of 
quantum correlations, Cirel'son's seminal work 
has been extended in a number of ways. 
The specific case of the CHSH inequality has been further studied in 
\cite{L,BMR,PR,CB,D}. In the case of two settings and two outcomes at 
each site all 
Cirel'son bounds are known for inequalities of the CHSH type 
(correlation inequalities) \cite{C2}, but not in the general 
case. Very little is known when the 
number of settings and the number of outcomes increases 
(see however the numerical approach of 
\cite{FS}), 
although the quantum correlations appear to have 
a very interesting structure already for dimension 3, see for instance 
\cite{CGLMP,Gd=3}. A 
very few results are also known in the case of more than 2 
parties\cite{Cmulti}. 

Thus there clearly remains a lot to be learned about the boundary of the 
space of correlations obtainable by local measurements on entangled 
quantum states. Better understanding this boundary is of fundamental 
interest as mentioned above, but is also of practical interest as 
understanding this boundary may enable the derivation of 
better experimental tests of quantum non locality, and may suggest 
better ways in which 
quantum 
non locality can help in information processing tasks such as 
communication complexity\cite{BC}.

All these earlier works on Cirel'son's bound proceeded essentially in 
the same way, namely they first imposed that the measurements carried 
out on the entangled states are local, and then tried to find, given 
this condition, what is the maximal value attainable by a Bell 
inequality. Here we shall adopt another approach which at first sight may seem 
surprising, even paradoxical, but which is very much in the spirity of 
the work of Popescu and Rohrlich. Indeed 
we shall explore what are the consequences of requiring the 
coexistence of the following apparently contradictory conditions: 1) 
the existence of correlations that obey the no signalling condition 
and which violate Cirel'son's bound; 2) the validity of quantum 
mechanics. The only way to make these apparently contradictory requirements 
compatible is to drop the locality condition: it must be that during 
the measurement process some communication took place between the 
parties. Hence one is dealing with a single global, rather than 
several local, measurements. 

The communication that takes place during the measurement 
is however ``hidden''. Indeed the correlations obey 
the no signalling condition. Thus although communication was necessary 
to produce the correlations, the correlations themselves cannot be used to 
communicate. 
The 
question of finding Cirel'son bounds can thus be reformulated as the 
question of finding whether or not, within the framework of quantum 
mechanics, 
such ``hidden'' communication is necessary to 
obtain the correlations.

Our first result is to show the (surprising) fact that, 
if the whole measurement process 
including the hidden communication which takes place during the 
measurement 
is a unitary reversible process, then this hidden 
communication can always be "revealed". 
By this we mean that if the measurement process is implemented by some unitary evolution $U$, then by carrying out local quantum operations the parties can in fact signal and generate entanglement. More specifically this is realised by
carrying out a coherent superposition of different measurements. 
This result is completely general. In fact it characterises the boundary of the space of correlations 
obtainable by carrying out local measurements on entangled quantum 
states. This first result is closely related to earlier work devoted characterising the 
the constraints that locality and causality impose on quantum operations, see\cite{BGNP,BHLS,PHHH}.
It is also closely related to -but more general than- 
work of Dieks\cite{D} who 
also emphasized the causality conditions implicit in derivations of 
Cirel'son bounds.

Because of its generality one can turn the above remark into a tool to 
study Cirel'son bounds. This is done as follows: take some 
correlations and suppose that the quantum measurement process that 
gave rise to them is unitary and reversible. Let the parties carry out 
a coherent superposition of different measurements and try to show that such a 
superposition measurement generates entanglement, or allows 
communication. If you succeed you have shown that these correlations 
exceed Cirel'son's bound. Thus whereas previous studies of 
Cirel'son bounds approached the bound from below by exibiting 
correlations that could be attained using local quantum measurements, 
here we have a method of approaching the bounds from above, by 
showing that some correlations cannot be obtained using local 
measurements on entangled quantum states. The two approaches are thus 
complementary.

We apply this method to several examples. First we sudy the simple 
example of the correlations studied by Popescu and Rohrlich that 
maximally violate the CHSH inequality and show how in this case the unitary $U$ allows signalling and entanglement generation. We also generalise this example to the case where the number of measurement settings and the number of measurement outcomes is $d$. 
Then we rederive Cirel'son's 
bound for the CHSH inequality. Finally we study a generalisation of 
the CHSH inequality involving 3 measurement settings for each party 
and 3 outcomes for each measurement. We derive a Cirel'son bound in 
this case although, to our knowldge, none was known previously in this case. 
We also point out an important discussion at the end of section \ref{charact} devoted 
to the interpretation of our 
approach, and in particular its relation to the program initiated by 
Popescu and Rohrlich whose aim is to understand why the quantum 
correlations are not maximally non local.

\section{Non Local Correlations}\label{nonlocalcorr} 

Consider the following situation: there are two parties, Alice and 
Bob, who each choose an input (their measurement setting), x 
and y, and produce an output (their measurement outcome), a and b. 
Upon repeating this 
experiment many times one can describe it synthetically by the set of 
probability distributions $P(a,b|x,y)$ (the probabilities of the 
outputs conditional on the inputs). 
We impose that the correlations obey the no-signalling conditions 
\begin{eqnarray} 
&\sum_{a}P(a,b|x,y)=P(b|y)\mbox{ is independent of $x$}&\nonumber\\ 
&\sum_{b}P(a,b|x,y)=P(a|x)\mbox{ is independent of $y$} 
\label{Nsig} 
\end{eqnarray} 
which express the fact 
that Alice's(Bob's) output cannot 
provide 
her(him) 
with any information about Bob's(Alice's) input. 

Of course if Alice and Bob's measurements, starting from the choice of 
inputs $x$ or $y$ and ending with the production of the output $a$ or 
$b$, take place in spatially separated locations, then the no 
signalling conditions must be obeyed. But as mentioned in the 
introduction we shall also consider correlations which require some 
communication between Alice and Bob. 
If the no signalling conditions are 
not obeyed then it is immediatly obvious that 
communication is required to obtain the correlations $P(a,b|x,y)$ and 
the whole question of Bell inequalities and Cirel'son bounds becomes 
irrelevant. Thus we always impose the no signalling conditions.

Let us suppose that the above measurement process can be described 
entirely within the context of the 
quantum formalism. Without loss of generality we can describe it as 
follows: initially the parties start with the state 
\begin{equation} 
|\Psi^0_{xy}\rangle = |x\rangle_A |y\rangle_B |0\rangle_A |0\rangle_B 
|\psi\rangle_{AB} \label{Psi0} 
\end{equation} 
where $|x\rangle_A |y\rangle_B$ are the quantum states containing 
Alice and Bob's inputs (ie. they specify the measurement settings), $|0\rangle_A |0\rangle_B$ are the initial 
states of Alice and Bob's output registers, as well as the initial 
state of any local ancillas they may use, $|\psi\rangle_{AB}$ 
is the entangled state Alice and Bob share. The subscripts indicate to 
which subspace, Alice's, Bob's, or both, the states belong. 

In order to produce their outputs, Alice and Bob carry out some 
unitary operation $U$ to obtain the state: 
\begin{eqnarray} 
|\Psi^1_{xy}\rangle =U |\Psi^0_{xy}\rangle 
= \sum_{a,b}|a\rangle_A |b\rangle_B 
|\varphi^{ab}_{xy}\rangle_{AB}\ . \label{Psi1} 
\end{eqnarray} 
Here $|a \rangle_A$ and $|b\rangle_B$ are the final states of Alice and Bob's 
output registers. 
If Alice and Bob carry out local measurements, then $U= U_A \otimes 
U_B$ must have a product form. If Alice and Bob carry out some 
communication during the measurement, then $U$ will in general not be 
a product.

In all cases the unormalised states $ |\varphi^{ab}_{xy}\rangle_{AB}$ 
satisfy the orthogonality relations 
\begin{equation} 
\sum_{a,b} \langle \varphi^{ab}_{xy}| \varphi^{ab}_{x'y'}\rangle = 
\delta_{x x'} 
\delta_{y y'} 
\label{orth} 
\end{equation} 
Using the above notation, the probabilities of finding outcomes $a$ and $b$ 
conditional on inputs $x$ and $y$ is 
\begin{equation} 
P(a,b|x,y)=\langle \varphi^{ab}_{xy}|\varphi^{ab}_{xy}\rangle\ . 
\label{PA} 
\end{equation} 

The space of correlations $P(a,b|x,y)$ which can be obtained using 
local measurements, ie. using product operations $U=U_A\otimes U_B$, 
is easily shown to be a convex set. Hence it can be characterised by 
linear inequalities 
\begin{equation} 
{\bf b}\cdot {\bf P} = \sum_{a,b,x,y} b_{a,b,x,y}P(a,b|x,y) \leq 
b_{QM} \ . 
\label{Cir} 
\end{equation} 
Here $b_{QM}$ is the largest value of ${\bf b}\cdot {\bf P}$ 
obtainable using local measurements: 
\begin{equation} 
b_{QM} = \max_{\psi_{AB}, U=U_A\otimes U_B} 
\sum_{x,y,a,b}b_{abxy}\langle 
\varphi_{xy}^{ab}|\varphi_{xy}^{ab}\rangle\ . 
\label{BQM}\end{equation} 
The set of inequalities ${\bf b}\cdot {\bf P} \leq 
b_{QM}$ consititute the Cirels'on bounds. 

\section{Characterising Non Local Correlations in Terms of Signalling} 
\label{charact} 

Let us now prove our first result. We consider some correlations 
$P(a,b|x,y)$ that obey the no signalling condition. These 
correlations must 
satisfy positivity 
$P(a,b|x,y)\geq 0$ and normalization $\sum_{a,b}P(a,b|x,y)=1$. Hence 
there exists some transformation of the form 
(\ref{Psi0})$\to$(\ref{Psi1}), realised by some $U$, that implements 
the measurements leading to these correlations. We will show that if 
these correlations violate a Cirel'son 
bound of the form eq. (\ref{Cir}), then the transformation $U$ 
can be used to generate entanglement and to 
carry out classical communication. This result holds for all values of 
the number of measurement settings and measurement outcomes. 

To prove this we reason as follows. 
In view of definition (\ref{BQM}), $U$ cannot be 
written as a tensor product: $U\neq U_A\otimes U_B$. 
However it is easy to show that any unitary that cannot be written as 
a tensor product can generate entanglement. The idea of the argument 
is to consider the action of 
$U_{AB}$ on the maximally entangled state $|\psi\rangle_{A'ABB'}=N \sum_i 
|i\rangle_{A'}|i\rangle_A\sum_j|j\rangle_B |j\rangle_{B'}$ where $N$ is 
a normalisation constant. It is easy to see that 
the state $|\psi^f\rangle_{A'ABB'}= I_{A'}\otimes U_{AB} 
\otimes I_{B'} |\psi\rangle_{A'ABB'}$ completely characterizes the 
transformation 
$U_{AB}$. If $U_{AB}=U_A\otimes U_B$, then there is no entanglement 
between systems $A'A$ and $BB'$ in state $|\psi^f\rangle$. 
Conversely if state 
$|\psi^f\rangle$ contains no entanglement between systems $A'A'$ and $BB'$, 
then the transformation $U_{AB}$ associated to $\psi_f$ is a tensor 
product. Hence separability of $\psi^f$ and tensor product character 
of $U_{AB}$ are equivalent statements. This implies that any unitary 
which cannot be written as a tensor product 
can generate entanglement by acting with it on the maximally entangled 
state $|\psi\rangle_{A'ABB'}$. We now refer to 
recent work of Bennett et al \cite{BHLS} in which it is shown that any 
unitary that can generate entanglement 
necessarily also allows signaling. This concludes the proof. 

Note that for the above proof to hold it is essential to allow the parties 
to manipulate coherently the registers, $|x\rangle$ and $|y\rangle$, which 
specify which measurement is going to be realised. Specifically they 
must be able to prepare superpositions of different register states, 
and to locally entangle these registers with other systems. We shall 
see this at work in the examples considered below.

At this point there is an important remark to make. The above argument 
supposed that there is no decoherence. Decoherence can of course be 
included in the description of section \ref{nonlocalcorr}: one adds to 
the 
description of the initial state eq. (\ref{Psi0}) the initial state of 
the environment; during the evolution the state of the measuring 
devices can get entangled with the environment; and at the end of the 
evolution one traces over the environment. But if there is such an 
environment which is inaccessible to the parties, it may be that the 
entanglement which necessarily exists if the correlations violate a 
Cirels'on bound, is in fact localised in the environment, and hence is 
inaccessible. Thus our first result only holds if no decoherence 
occurs, ie. if the evolution is reversible. 

This remark does not 
change anything to the applications we consider in sections \ref{SPC} and \ref{NC}, 
but it is an important remark from the point of view of interpretation. 
We suspect that reversibility plays a 
crucial role in understanding Cirel'son's bound. Indeed one 
can rephrase the program initiated by Popescu and Rohrlich as: ``What 
is the minimum set of axioms which imply Cirel'son's bound?''. Popescu 
and Rohrlich showed that causality is not enough. However adding 
reversibility may constrain the correlations much more. This idea is 
supported by the fact that Hardy recently proposed 
a set of axioms which imply 
quantum mechanics, and reversibility is the crucial one that 
differentiates classical from quantum mechanics\cite{H}. 
Our result above seems to suggest that both causality and reversibility could 
be enough to rule out the extremal correlations of Popescu and 
Rohrlich, and maybe even recover Cirel'son's bound. We do not know how 
to prove this, but we feel it is a very interesting avenue of 
research.

\section{Signaling with Perfect Correlations} 
\label{SPC} 

As discussed in \cite{C2,MNL}, for a fixed number of measurement settings 
and outcomes, the correlations which obey the no 
signalling conditions eq. (\ref{Nsig}) constitute a polytope of which the 
quantum correlations (those obtained using only entanglement and local 
quantum measurements) and classical correlations (those obtained using only 
local hidden variable models) are proper subsets. The most non local 
correlations are the extremal points, ie. the vertices, of the no 
signalling polytope. 

The correlations considered by Popescu and 
Rohrlich are extremal no signalling correlations in the above 
sense. Explicitly, if we suppose that $a,b,x,y\in\{0,1\}$, 
they take the form 
$P^{PR} (a,b|x,y)=1/2$ if $a\oplus b = xy$, $P^{PR} (a,b|x,y) = 0$ 
otherwise. 

In what follows we shall consider a generalisation of the 
Popescu-Rohrlich correlations to the case where the number of 
measurement settings is $d$ and the number of measurement outcomes is $d$. 
Specifically take $a,b,x,y\in \{0,\ldots,d-1\}$, then these 
correlations take the form 
\begin{equation} 
P(a,b|x,y)= \left\{ 
\begin{array}{lcr} 
1/d & \mbox{if} & a - b = xy {\mbox{ mod }} d\\ 
0 & \mbox{otherwise} & 
\end{array} 
\right. 
\label{P}\end{equation} 
(in the condition that appears in eq. (\ref{P}) the addition and 
multiplication are modulo $d$). 
We note that the correlations eq. (\ref{P}) are extremal points of the set of no signalling correlations with $a,b,x,y\in \{0,\ldots,d-1\}$.
These correlations coincide with the Popescu-Rohrlich 
correlations for $d=2$. 
(Note that this is not the only generalisation of the Popescu-Rohrlich 
correlations to higher dimensions. For instance in \cite{MNL} a 
generalisation to the case $x,y\in\{0,1\}, a,b\in\{ 0,...,d-1 \}$ was 
considered. Our method seems more difficult to apply in this case, and 
we therefore consider here the correlations 
eq. (\ref{P}).)

The argument we now present is loosely inspired by \cite{CvDNT}. 
Eq. (\ref{P}) implies that in eq. (\ref{Psi1}) the sum over 
$a,b$ is restricted to $a - b = xy {\mbox{ mod }} d $. 
Let us now show that in this case the unitary transformation $U$, that 
implements the correlations from eq. (\ref{P}, allows 
signaling and generation of entanglement. 
To this end suppose that after realising $U$ the parties carry out 
the local unitary transformations $e^{i 2\pi \hat a /d}$ and $e^{-i 2 \pi 
\hat b /d}$ where 
$e^{i 2 \pi \hat a /d} |a\rangle_A = e^{i 2 \pi a/d}|a\rangle_A$, 
$e^{-i 2 \pi \hat b /d} |b\rangle_B = e^{-i 2 \pi b/d} 
|b\rangle_B$. Because $a - b = xy {\mbox{ mod }} d$, the 
resulting state can 
be written 
as 
$$ 
e^{i 2\pi (\hat a - \hat b)/d} U |\Psi^0\rangle 
= e^{i 2 \pi xy/d} \sum_{a,b \atop a - b = xy \ mod\ d} 
|a\rangle_A |b\rangle_B 
|\varphi^{ab}_{xy}\rangle_{AB}\ . 
$$ 
Finally the parties carry out the inverse of the transformation $U$ to 
obtain 
\begin{eqnarray} 
|\Psi^f_{xy}\rangle&=& 
U^\dagger e^{i 2\pi (\hat a - \hat b)/d} U |\Psi^0\rangle 
\nonumber\\ 
&=& e^{i 2 \pi xy/d} |x\rangle_A |y\rangle_B 
|0\rangle_A |0\rangle_B 
|\psi\rangle_{AB}\ . 
\end{eqnarray} 
They have recovered the original state up to the phase 
$ e^{i 2 \pi xy/d}$. 

To signal the parties carry out the above series of transformations on 
a superposition of inputs. 
Indeed if Bob uses a uniform 
superposition of input states the evolution is 
\begin{eqnarray} 
&{1 \over \sqrt{d}} \sum_y | \Psi^f_{xy}\rangle_B 
=&\nonumber\\ 
& |x\rangle_A 
\left( {1 \over \sqrt{d}} \sum_y e^{i 2 \pi xy/d} |y\rangle_B \right) 
|0\rangle_A |0\rangle_B 
|\psi\rangle_{AB}\ .& 
\end{eqnarray} 
The states $|z\rangle = {1 \over \sqrt{d}} \sum_y e^{i 2 \pi zy/d} 
|y\rangle_B$ 
are orthogonal (they are the Fourier transform of $|y\rangle_B$). 
Hence by measuring his input variable in this basis Bob can 
learn Alice's input, ie. Alice has transmitted $\log_2 d$ bits of 
information to Bob.

One can show in a similar way that the transformation $U$ also allows 
generation of entanglement. Indeed 
if both parties use a coherent superposition of input 
states the final state is 
\begin{eqnarray} 
{1 \over d} \sum_{x,y} |\Psi^f_{xy}\rangle_A 
= {1 \over d} \sum_{x,y} e^{i 2 \pi xy/d} |x\rangle_A 
|y\rangle_B 
|0\rangle_A |0\rangle_B 
|\psi\rangle_{AB}\ . \label{7} 
\end{eqnarray} 
One easily checks that the final state contains $\log d$ ebits 
more entanglement than the initial state.

In summary by combining the transformation $U$, its inverse 
$U^\dagger$, and the local 
transformation $ e^{i 2\pi \hat a /d}$ and $e^{-i 2 \pi \hat b/d} $, 
the parties can either transmit $\log d$ bits of classical 
information or can generate $\log d$ ebit of entanglement. 
This would be impossible if $U$ was local and therefore shows that the correlations eq. (\ref{P}) violate a Cirel'son bound.

\section{Noisy Correlations and Cirel'son Bounds} 
\label{NC} 

The above example illustrates our main result in the case of extremal, 
ie. maximally non local, 
correlations. We now generalise it to the case where the correlations 
are noisy. We will suppose that the correlations are noisy versions of 
the extremal correlations 
eq. (\ref{P}). Our aim is to show that if the amount of noise is too 
small, then the unitary transformation $U$ that implements the 
measurement is non local and in particular allows the parties to 
communicate. 
In what follows we will rederive 
Cirel'son 
bound of $2\sqrt{2}$ for the CHSH inequality 
and also obtain a Cirel'son bound for correlations of the form 
eq. (\ref{P}) in the case $d=3$. 

In order to measure the degree of non locality of quantum correlations 
we will 
use a Bell expression of the form eq. (\ref{Cir}). We write it as: 
$B^d({\bf P})= {\bf b^d}\cdot {\bf P}$ with 
\begin{eqnarray} 
b_{abxy}^d= \left\{ 
\begin{array}{lcr} 
1/d^2 & \mbox{if}& a- b=xy {\mbox{ mod }} d \ ,\\ 
0 & \mbox{otherwise .} & 
\end{array} 
\right. 
\end{eqnarray} 
This Bell expression 
$B^d$ gives the probability, averaged over the inputs, that the 
outputs satisfy 
the relation $a-b=xy {\mbox{ mod }} d $, hence $0\leq B^d \leq 1$. 
Note that $B^2$ is related 
to the CHSH 
expression by the rescaling $B^2 = B^{CHSH}/8 +1/2$. 

The reason for 
using $B^d$ 
is that it allows us to compare the different values of $d$, since it 
is always bounded between 0 and 1. 
We will denote by $B^d(lhv)$ the values of $B^d$ obtainable in a local hidden variable model and by $B^d(QM)$ the values of $B^d$ obtainable using local measurements on entangled quantum states. In the case 
$d=2$ one has $B^2(lhv)\leq 3 / 4 =0.75$ and $B^2(QM) \leq 1 / 2 + 1/ 
2 \sqrt{2} 
\simeq 0.853$ (this is Cirel'son bound -rescaled-, which we rederive 
below). 
In the case 
$d=3$ it is not difficult to show, by enumerating all deterministic 
local classical strategies, that $B^3(lhv)\leq 2 / 3\simeq 0.66$ 
and we show below 
that $B^3 (QM) \leq 1/3 + 2/3\sqrt{3}\simeq 0.72$ (although we do not know 
whether this bound can be attained). This shows that as $d$ increases 
from 2 to 3 it is 
increasingly difficult to satisfy the relation $a-b=xy {\mbox{ mod }} 
d$, both 
classically and quantum mechanically. 
We have so far not been able to generalize these 
results to higher values of $d$. 

Our argument proceeds essentially as above. We suppose the parties start with 
state $|\psi^0_{xy}\rangle$ given in eq. (\ref{Psi0}). The main new ingredient is that 
before applying $U$ 
the parties 
first copy their input into a separate register on which $U$ does not 
act (this 
will provide us with a useful orthogonality relation later).
This is 
done using the unitary transformation which in the 
computational basis realises the copy operation. 
That is Alice 
carries out the local unitary operation $U^{copy}_{A}$ 
which acts as $U^{copy}_A |0\rangle^{copy}_A |x\rangle_A = 
|x\rangle^{copy}_A |x\rangle_A$, where $|0\rangle^{copy}_A$ is the 
initial state of the copy system. (Note that this operation only copies 
perfectly in the computational basis, it does not copy superpositions 
perfectly, hence it is not in contradiction with the quantum 
no-cloning theorem). 
Similary Bob carries out $U^{copy}_{B}$ 
which acts as $U^{copy}_B |0\rangle^{copy}_B |y\rangle_B = 
|y\rangle^{copy}_B |y\rangle_B$. 

Thus we have 
\begin{eqnarray} 
&U^{copy}_A\otimes U^{copy}_B |0\rangle^{copy}_B |0\rangle^{copy}_A 
|\Psi^0_{xy}\rangle&\nonumber\\ 
& = |x\rangle_A^{copy}|y\rangle_B^{copy} 
|x\rangle_A|y\rangle_B |\psi\rangle_{AB}\ .& 
\label{x}\end{eqnarray} 
The parties now act with $U$ (We suppose that $U$ does not act on the copied 
inputs) to yield 
\begin{eqnarray} 
&U U^{copy}_A\otimes U^{copy}_B |0\rangle^{copy}_B |0\rangle^{copy}_A 
|\Psi^0_{xy}\rangle=& 
\nonumber\\ 
&|x\rangle_A^{copy}|y\rangle_B^{copy} 
\sum_{a,b} |a\rangle_A|b\rangle_B |\varphi^{ab}_{xy}\rangle_{AB}\ .& 
\label{UU}\end{eqnarray} 
Since $U$ does not perfectly reproduce the correlations 
eq. (\ref{P}) 
all values of $a,b$ can appear in eq. (\ref{UU}). For simplicity of 
notation we 
shall denote hereafter 
\begin{equation} 
|\varphi^k_{xy}\rangle = \sum_{a,b \atop a-b-xy=k \ mod\ d} 
|a\rangle_A|b\rangle_B |\varphi^{ab}_{xy}\rangle_{AB}\ . 
\label{phik} 
\end{equation} 

The parties now carry out the local operations 
$e^{i 2 \pi \hat a/d}\otimes e^{-i 2 \pi \hat b/d}$ to obtain 
\begin{eqnarray} 
& 
e^{i 2 \pi \hat a/d}\otimes e^{-i 2 \pi \hat b/d} 
U 
U^{copy}_A\otimes U^{copy}_B |0\rangle^{copy}_B |0\rangle^{copy}_A 
|\Psi^0_{xy}\rangle 
=& 
\nonumber\\ 
& e^{i 2 \pi xy 
/d}|x\rangle_A^{copy}|y\rangle_B^{copy} 
\sum_{k} e^{i 2 \pi k/d}|\varphi^{k}_{xy}\rangle\ . 
\label{UU'}&\end{eqnarray} 

Finally the parties carry out the inverse operations $U^{\dagger}$ and 
$U^{copy \dagger}_A\otimes U^{copy \dagger}_B$ to obtain 
\begin{eqnarray} 
|\Psi^f_{xy}\rangle 
&=& U^{copy \dagger}_A\otimes U^{copy \dagger}_B 
U^{\dagger}e^{i 2 \pi \hat a/d}\otimes e^{-i 2 \pi \hat b/d} \nonumber\\ 
& & 
U 
U^{copy}_A\otimes U^{copy}_B |0\rangle^{copy}_B |0\rangle^{copy}_A 
|\Psi^0_{xy} \rangle
\ . 
\end{eqnarray} 

Our aim is to find the maximal value of the Bell expression 
$\langle B^d \rangle = {1 \over d^2}\sum_{x,y,a,b \atop a-b-xy=0 \ mod\ d} 
P(a,b|x,y) 
= 
{1 \over d^2}\sum_{x,y} \langle 
\varphi^0_{xy}|\varphi^0_{xy}\rangle$ for which the transformation 
$U$ can be implemented without communication, ie. for which 
$U=U_A\otimes U_B$ is local. To this end we shall investigate for 
what values of $B^d$ the transformation 
$ 
U^{copy \dagger}_A\otimes U^{copy \dagger}_B 
U^{\dagger} e^{i 2 \pi \hat a /d}\otimes e^{-i 2 \pi \hat b/d} U 
U^{copy}_A\otimes U^{copy}_B$ 
does not 
necessarily imply that Alice and Bob can signal. 
We suppose that initially Bob prepares his input in a coherent 
superposition ${1\over \sqrt{d}}\sum_{y} |y \rangle_B$ whereas Alice prepares 
her input in state $|x\rangle_B$. 
After carrying out the operations 
$ 
U^{copy \dagger}_A\otimes U^{copy \dagger}_B 
U^{\dagger} e^{i 2 \pi \hat a /d}\otimes e^{i 2 \pi \hat b/d} U 
U^{copy}_A\otimes U^{copy}_B$ 
Bob measures his 
input in the basis $|z\rangle_B = {1\over \sqrt{d}} \sum_{y} 
e^{i 2 \pi zy/d} |y 
\rangle_B$. The probability of finding $z$ given that Alice prepared $x$ is 
\begin{equation} 
P(z|x)=\left( {1\over \sqrt{d}}\sum_{y'} \langle\Psi^f_{xy'}| \right) 
\Pi_z 
\left( 
{1\over 
\sqrt{d}}\sum_{y} |\Psi^f_{xy}\rangle \right) 
\end{equation} 
where $\Pi_z= |z\rangle_{B}\langle z |$ projects onto state 
$|z\rangle_B$ and acts as the identity on the rest of the Hilbert space. 
We use the fact that 
\begin{eqnarray} 
\Pi_z &\geq& \left({1\over \sqrt{d}}\sum_{y'} e^{-i 2 \pi zy'/d} 
|0\rangle^{copy}_B |0\rangle^{copy}_A |\Psi^0_{xy'}\rangle\right)\nonumber\\ 
& & 
\left({1\over \sqrt{d}}\sum_{y}e^{i2 \pi yz/d} 
\langle0|^{copy}_B \langle0|^{copy}_A\langle\Psi^0_{xy}| \right) 
\nonumber 
\end{eqnarray} 
where the $\geq$ sign means that the difference of the two operators is a positive operator.
This implies that 
\begin{eqnarray} 
\!\!\!\!\!\!\!\!\! 
P(z|x)&\geq& | {1\over d}\sum_{yy'} e^{-i 2 \pi zy'/d} 
\langle0|^{copy}_B \langle0|^{copy}_A\langle\Psi^0_{xy'}| 
\nonumber\\ 
& & 
\!\!\!\!\! 
U^{copy \dagger} U^{\dagger} e^{i 2 \pi (\hat a -\hat b)/d} U U^{copy} 
|0\rangle^{copy}_B |0\rangle^{copy}_A|\Psi^0_{xy}\rangle |^2\nonumber\\ 
&=&|{1\over d}\sum_{y} e^{i 2 \pi y(x-z)/d} \sum_k e^{i 2 \pi k/d} \langle 
\varphi^k_{xy}|\varphi^k_{xy}\rangle|^2 
\label{PPP} 
\end{eqnarray} 
where 
we denote $U^{copy}=U^{copy}_A\otimes U^{copy}_B$ and 
we have used eq. (\ref{x}) and 
$_B^{copy}\langle 
y'|y\rangle^{copy}_B = \delta_{y' y}$. 
The reason for introducing 
the copy operations $U^{copy}_{A(B)}$ is now clear: without them 
we would have obtained a similar 
expression to eq. (\ref{PPP}), but with a double sum over $y$ and $y'$ rather 
than the single sum over $y$. 

If the transformation $U$ is no signaling, it must be that Bob cannot learn 
what Alice's input was. 
This implies that $P(z|x)=P(z|x')$ for all $z,x',x$. We 
now combine this with the normalization condition $\sum_{z}P(z|x)=1$ to obtain 
the condition 
$$\sum_x P(x|x)=1\ .$$ 
Replacing $P(x|x)$ in this equality by 
the bound eq. (\ref{PPP}) yields the inequality 
\begin{equation} 
1 \geq \sum_x |{1\over d}\sum_{y} \sum_k e^{i 2 \pi k/d} \langle 
\varphi^k_{xy}|\varphi^k_{xy}\rangle|^2\ . 
\label{PPPP} 
\end{equation} 
We now use that if $\sum_{j=1}^d |x_j|^2 \leq 1$, then 
$\sum_{j=1}^d |x_j| \leq 
\sqrt{d}$, to obtain 
\begin{equation} 
\sqrt{d} \geq \sum_x|{1\over d}\sum_{y} \sum_k e^{i 2 \pi k/d} \langle 
\varphi^k_{xy}|\varphi^k_{xy}\rangle|\ . 
\label{P5} 
\end{equation} 
Using eqs. (\ref{PA}) and (\ref{phik}) we can rewrite this as 
\begin{equation} 
\sqrt{d} \geq \sum_x|{1\over d}\sum_{y} \sum_k e^{i 2 \pi k/d} 
\sum_{a,b \atop a-b-xy=k \ mod\ d} P(a,b|xy)|\ . 
\label{C1} 
\end{equation} 
This is a Cirel'son inequality: any correlations that violate it 
cannot be reproduced using local measurements on entangled quantum 
states. We will now show that the inequality (\ref{C1}) is 
tight in the case of $d=2$: it is equivalent to the bound $2\sqrt{2}$ 
on the CHSH inequality. We will also show that in the case $d=3$ it 
gives a non trivial bound on $B^3$, as anounced above. 
However in the case $d=4$ it apparently does not give an interesting bound, 
presumably 
because it does not incorporate enough of the no signaling constraints.

First we simplify eq. (\ref{C1}) by using $|x|\geq |Re(x)|\geq Re(x)$ to obtain 
\begin{equation} 
\sqrt{d} \geq \sum_x{1\over d}\sum_{y} \sum_k \cos (2 \pi k/d) 
\sum_{a,b \atop a-b-xy=k \ mod\ d} P(a,b|xy)\ . 
\label{C2} 
\end{equation} 

We now specialize to the case $d=2$. 
Eq. (\ref{C2}) then becomes 
\begin{eqnarray} 
2 \sqrt{2} &\geq& \sum_x \sum_{y} \left( 
\sum_{a,b \atop a-b-xy=0 \ mod\ 2} P(a,b|xy) 
\right. 
\nonumber\\ 
&& \left. 
-\sum_{a,b \atop a-b-xy=1 \ mod\ 2} P(a,b|xy) \right) 
\ . 
\label{C3} 
\end{eqnarray} 
which is Cirel'son's bound for the CHSH expression. 

In the case $d=3$, eq. (\ref{C2}) becomes 
\begin{eqnarray} 
3 \sqrt{3} &\geq& \sum_{x=0}^2 \sum_{y=0}^2 \left( 
\sum_{a,b \atop a-b-xy=0 \ mod\ 3} P(a,b|xy) 
\right. 
\nonumber\\ 
&& 
-{1\over 2}\sum_{a,b \atop a-b-xy=1 \ mod\ 3} P(a,b|xy) 
\nonumber\\ 
&& \left. 
-{1\over 2}\sum_{a,b \atop a-b-xy=2 \ mod\ 3} P(a,b|xy) 
\right) 
\ . 
\label{C4} 
\end{eqnarray} 
We now use that $\sum_{a,b}P(a,b|xy)=1$ to eliminate the last two terms and 
obtain 
\begin{equation} 
{9\over 2} + 3 \sqrt{3} \geq {3\over 2} \sum_{x=0}^2 \sum_{y=0}^2 
\sum_{a,b \atop a-b-xy=0 \ mod\ 3} P(a,b|xy) 
\ . 
\label{C5} 
\end{equation} 
Upon multiplication of eq. (\ref{C5}) by $2/3^3$ one obtains $B^3(QM)\leq {1 \over 3}+ {2 \over 3 \sqrt{3}}$ as announced.

\section{Conclusion} 
In the present work we have studied the properties of no signalling 
correlations that cannot 
be reproduced using local measurements on entangled quantum 
states. Such correlation we say violate Cirel'son bounds. We 
supposed that the correlations are obtained by some  reversible unitary
quantum evolution $U$. Because the correlations violate a Cirel'son 
bound the evolution $U$ cannot be written in the product form $U_A 
\otimes U_B$. We show that this implies that $U$ can be used for 
signalling and for entanglement generation. This approach is very 
general and in fact can be viewed as a complete characterisation of 
Cirel'son bounds. We show that it can be used as a tool to study 
Cirel'son bounds and we illustrate this by rederiving the Cirel'son bound 
of $2\sqrt{2}$ for the CHSH inequality, and by deriving a new 
Cirel'son bound for qutrits.

{\bf Acknowledgements:} We acknowledge financial support by project 
RESQ IST-2001-37559 of the IST-FET program of the EC, by the 
Communaut\'e Fran{\c {c}}aise de Belgique under grant ARC00/05-251, 
the IAP program of the Belgian governement under grant V-18, and by 
the Netherlands Organisation for Scientific Research (NWO) VICI grant 
nr: 639.023.302.


\begin{thebibliography}{99} 

\bibitem{B} J. S. Bell, Physics {\bf 1} (1964) 195 

\bibitem{CHSH}J. F. Clauser, M. A. Horne, A. Shimony, R. A. Holt, 
Phys. Rev. Lett. 23 (1969) 880 

\bibitem{C} B. S. Cirel'son, Lett. Math. Phys. {\bf 4} 93 (1980) 

\bibitem{PR} S. Popescu and D. Rohrlich, Found. Phys. {\bf 24} 379 
(1994), see also quant-ph/9709026

\bibitem{C2} B. S. Cirel'son, Hadronic Journal Supplement, {\bf 8} 
(1993) 329 

\bibitem{MNL} J. Barrett, N. Linden, S. Massar, S. Pironio, 
S. Popescu, D. Roberts, Phys. Rev. A 71, 022101 (2005) 


\bibitem{L} L. J. Landau, Phys. Lett. A {\bf 120} (1987) 54 

\bibitem{BMR} S. L. Braunstein, A. Mann, M. Revzen, Phys. Rev. Lett. {\bf 68} 
(1992) 3259 


\bibitem{CB} A. Chefles, S. M. Barnett, J. Phys. A {\bf 29} (1996) L237 


\bibitem{D}D. Dieks, Phys. Rev. A {\bf 66} 062104 (2002) 

\bibitem{FS} S. Filipp, K. Svozil, Phys. Rev. Lett. 93, 130407 (2004) 





\bibitem{CGLMP} D. Collins, N. Gisin, N. Linden, S. Massar, 
S. Popescu, Phys. Rev. Lett. {\bf 88} (2002) 040404 

\bibitem{Gd=3} A. Acin, T. Durt, N. Gisin, J. I. Latorre, Phys. Rev. A 
{\bf 65} (2002) 052325 



\bibitem{Cmulti} P. Mitchell, S. Popescu, D. Roberts, Phys. Rev. A 70, 060101(R) (2004) 

\bibitem{BC} R. Cleve, H. Buhrman, Phys. Rev. A {\bf 56} (1997) 1201 

\bibitem{BGNP} D. Beckman, D. Gottesman, M. A. Nielsen, J. Preskill, Phys. Rev. A {\bf 64} (2001) 052309 



\bibitem{BHLS} C. H. Bennett, A. W. Harrow, D. W. Leung, J. A. Smolin, 
IEEE Trans. Inf. Theory, Vol. 49, No. 8, (2003) 1895 

\bibitem{PHHH} M. Piani, M. Horodecki, P. Horodecki, R. Horodecki, quant-ph/0505110


\bibitem{CvDNT} R. Cleve, W. van Dam, M. Nielsen, A. Tapp, Lect. Notes 
Comput. Sci. 1509 (1998) 61 


\bibitem{H} L. Hardy, "Quantum theory from five reasonable axioms", quant-ph/0101012 










\end{thebibliography}
\end{document}